# The Economics of Climate Adaptation: An Assessment

Anna Josephson[1*], Rodrigo Guerra Su[1], Greg Collins[2], Katharine Jacobs[3]

## Abstract

The cost of the impacts of climate change have already proven to be larger than previously believed. Understanding the costs and benefits of adapting to the changing climate is necessary to make targeted and appropriate investment decisions. In this paper, we use a narrative review to synthesize the current literature on the economic case for climate adaptation, with the objective of assessing the value (economic and otherwise) of climate change adaptation, as well as the strength of the methods and evidence that have been used to date. We find that skepticism is warranted about many of the estimates about costs and benefits of climate adaptation and their underlying assumptions, due to a range of complexities associated with (1) uncertainty in distinguishing the economic impacts of climate change from seasonal variability; (2) difficulties in non-market valuation; (3) lack of consistent data collection over time at multiple scales; and (4) distributional inequities in access to proactive adaptation and recovery funding. While useful for broad stroke advocacy purposes, these estimates fall short of the refinement and rigor needed to inform investment decision-making, particularly at micro and local scales. Most estimates rely on cost benefit analysis and don't effectively address these issues. An emergent and promising literature tackles alternative estimation strategies and attempts to address some of them, including the complexities of uncertainty and non-market valuation.

This work received funding from the USAID Climate Adaptation Support Activity (CASA).
[*] Corresponding author: aljosephson@arizona.edu.
[1] Department of Agricultural and Resource Economics, University of Arizona
[2] Resilience and International Development, University of Arizona
[3] Center for Climate Adaptation Science and Solutions & Department of Environmental Science, University of Arizona

# 1. Introduction

Climate change adaptation encompasses processes and actions undertaken by individuals, households, municipalities, nations, and global actors to reduce vulnerability to climate impacts and to enhance the capability to capture any benefits of climate change within natural and human systems. Understanding the costs and benefits of anticipatory climate change adaptation is imperative, both because the cost of climate change itself has already proven to be larger than previously believed and because understanding the costs and benefits of adapting to the changing climate are necessary to make targeted and appropriate investment decisions. In this paper, we seek to synthesize and understand the current literature on the economic case for climate adaptation, with the objective of assessing the value (economic and otherwise) of climate change adaptation, as well as the strength of the methods and evidence behind it.

In this paper, we conduct a narrative structured review of the literature on the economics of climate adaptation. We focus on identifying (1) the case for investment in climate adaptation and climate resilience; and (2) the economic evidence that substantiates the value of well-directed adaptation investments in sustaining development investments and reducing impacts in the face of climate change, including evidence on both cost-benefits and avoided losses. To this end, we assess prominent published and gray literature on the economic case for adaptation.[1] We aim to unpack the methodological underpinnings and assumptions of the economic case for climate adaptation in the existing literature, and to provide our assessment of the findings and conclusions of seminal work in this field. We do not attempt to assess the credibility of the claims, but to contextualize and analyze them. Though we examine sector specific evidence, we generally avoid discussion of specific interventions within and across sectors. We focus instead on evidence related to overall integration of approaches to adapt to climate risk and change.

This paper contributes to the literature as a concise synthesis of the existing literature on the economic case for climate adaptation, and the methods underlying these estimates. Outside and beyond the academic community, this work also has the potential to influence policymakers and decision-makers in the space of climate adaptation investment, through elucidating the process of estimating the economic impacts of climate change and the potential benefits and cost of various investments in adaptation. The paper also describes critical gaps where additional research and novel approaches are needed to further substantiate the economic case for climate adaptation to inform investment decision-making.

The literature around climate change costs, climate adaptation, and climate resilience is broad and expansive. To set boundaries around the universe of potentially available literature and documents, and in line with methods of structured and systematic reviews, we do not consider all the available literature in the adaptation/climate resilience space. Instead, our search method is grounded in a set of foundational documents, emphasizing the prominent evidence therein and extending and expanding from these seminal works. This review proceeds as follows: we discuss the method for searching citations in Section 2; Section 3 reviews the findings from the literature; Section 4 presents limitations and considerations, based on the findings of the literature; and our conclusions are found in Section 5.

---

[1] As Watkiss (2015) notes, much of the literature on this topic is gray literature. While there is more evidence emerging in the peer-reviewed literature, this review concurs that much of the relevant literature exists within the unpublished, gray literature.

## 2. Method

We conducted a narrative, structured review of the evidence on the economic case for climate adaptation, exploring evidence on the value of well-directed adaptation investments in sustaining development investments and reducing impacts in the face of climate change. We investigate both the peer-reviewed and published, as well as the gray literature on the topic. Though we use methods from systematic reviews, we create a narrative review which draws stylized qualitative patterns from the results and findings from the included literature. By contrast, a traditional, systematic approach requires a quantitative analysis with uniformity in outcomes and measures. Given the literature, focus of review, and objective of this work, the former was deemed more appropriate than the latter. As such, we undertook methods used in similar narrative reviews, to leverage the diversity and breadth of insights in the literature.

The review is situated and based in a set of foundational literature presented in Table 1. Fourteen high-impact sources were selected to serve as the foundational documents. Based on these foundational documents, we conducted a forward citation search and a snowball search.

| Table 1: Preliminary Results of Forward Citation Search | | | |
|---|---|---|---|
| **Foundational Document** | **Citation Count** | **Citation Count since 2018** | **Relevant Literature in Citation Count since 2015** |
| Abidoye (2021) | 5 | 5 | 0 |
| AdaptNow (2019) | Not searchable on Google Scholar | | |
| Chambwere et al. (2014) | 254 | 163 | 3 |
| Downing (2012) | 89 | 31 | 0 |
| ECONADAPT (n.d.) | Not searchable on Google Scholar | | |
| Fankhauser & Soare (2013) | 35 | 8 | 1 |
| Hallegatte, Lecocq, & de Perthuis (2011) | 131 | 47 | 3 |
| Hallegatte (2012) | 187 | 113 | 2 |
| Sovacool, Linner, & Goodsite (2015) | 213 | 169 | 0 |

| Stern (2006) | 9,608 | 1,700 | 6* |
| --- | --- | --- | --- |
| Tunis Roundtable (2010) | 38 | 9 | 0 |
| USAID (2013) | Not searchable on Google Scholar | | |
| Watkiss (2015) | 32 | 23 | 3 |
| Watkis et al. (2015) | 156 | 107 | 14 |
| * Search ended on page 5, after two pages of non-relevant pages. Search continued through page 7, with no more articles screened as relevant. | | | |

The foundational documents and their reference lists provide the base for both initial searches: a snowball search and a forward citation search. First, in a snowball search, additional citations and literature are found, looking backwards from a limited set of sources. A researcher follows references, citations, and recommendations to obtain additional information, as relevant to the scope of work being undertaken. As such, the reference list in a snowball search grows larger as the research leads, moving from the limited set of sources to those referenced therein, and beyond.[2] Next, in a forward citation search, additional citations and literature are found, expanding forward in time, from a limited set of sources. A researcher examines lists of sources which have cited a paper from the preliminary set of sources. The purpose of this type of search is to understand how the literature has changed since the publication of the original set of sources and how they have influenced subsequent research.

For both search types, we used the foundational documents as the focused set of sources from which the searches begin. For the snowball search, we do not set a timeframe for consideration of references, though we focus primarily on papers published since 2000. For the forward search, we focus on literature from 2018 onwards. Preliminary findings from our forward citation search are presented in Table 1. As the table notes, several references were not searchable in Google Scholar and so we did not conduct a forward search for those. Additionally, for several sources the forward search is ongoing. This is noted in the final column "Relevant Literature in the Citation Count since 2018".

In addition to the 14 foundational documents, this review includes 98 papers which were found as part of our snowball and forward citation searches.

## 3. Findings

A view of any literature that relies heavily on modeling should call to mind statistician George Box's comment: "Remember that all models are wrong; the practical question is how wrong do they have to be to not be useful" (Box, 1976). The literature on the economics of climate adaptation is no exception and is perhaps emblematic of this adage. The fields of economics and climate change adaptation both rely heavily on assumptions rather than measurements and empirical evidence. The combination of the two are, unsurprisingly, doubly reliant on underlying assumptions in generating estimates about the economics of climate adaptation.

---

[2] A complete snowball search was conducted for all papers except for the ECONADAPT (2015) report, where a more limited snowball search was conducted. In this case citations were often missing, incorrectly formatted, or with inaccurate DOIs such that snowballing from the document ranged from difficult to impossible.

The World Bank (2010) states that given the challenges in understanding adaptation and the difficulties in estimating its costs, benefits, and value, valuation becomes little more than "sophisticated guesswork". However, as Stern (2006) notes: while the cost of adaptation actions may be costly and difficult, the delay is much more dangerous and potentially costly. So, even with only sophisticated guesswork, we must work within the literature and estimates that we have while simultaneously identifying gaps and ways to strengthen estimation methods and the evidence base for investment decision-making.

In this section, we review the current evidence in the literature, founded in our snowball and citation searches of the foundational documents, with consideration of the caveats referenced above. This is followed by a synthesis of the problems identified within the literature, including shortcomings and opportunities for future work. This includes the challenges of uncertainty, missing markets, distributional inequalities, and lack of data/monitoring and feedback loops to accurately and precisely measure the value of climate adaptation.

*Evidence in the Literature*

For decades, researchers have recommended that any analysis of adaptation view the process through an explicitly economic lens (Carter et al., 1994; Parry and Carter, 1998). In even the earliest studies following this framework the dominant conclusion in the literature is that the benefits of investing in anticipatory climate change adaptation are strong and cost-effective, and that early actions far outweigh the economic costs of not acting (Stern, 2006).[3] With this review, we use this understanding as a starting point, seeking to understand the comparative value of these investments, to quantify that value, and to assess and understand the methods and evidence used to arrive at the value. These elements are necessary in order for decision-makers to make difficult choices, when there are costs and tradeoffs associated with the decisions. Though we agree with many authors that investments in climate preparedness can dramatically reduce the economic and non-economic cost of climate impacts, this review has reinforced our perspective that the evidence is not well documented.

Despite decades of work identifying the importance of analysis of the economics of adaptation, reviews of the literature are critical of its state. Adger et al. (2007) write that the literature on adaptation generally is "quite limited and fragmented" while more recently Jenal (2019) concludes similarly that there is no systematic change framework, which might guide research in this area. Carter et al. (2007) observe that in spite of the calls to do so, little of the literature on adaptation explicitly considers economics. Agrawala and Fankhauser (2008) concur, suggesting that there is little quantifiable economic evidence in the adaptation literature and what there is should be regarded with skepticism. Chapagain et al. (2020) dig into this literature further and provide an interesting and in-depth examination and analysis of some of the most recent figures, though ultimately they conclude that there is no international scientific or political agreement on current and projected adaptation costs.

---

[3] As stated by Mendelsohn (2012), we follow the definition of an investment being cost-efficient when the marginal benefit of the climate change adaptation intervention exceeds its marginal costs.

An estimation of "topline" or "headline" figures is common throughout the literature, in particular for large global or regional adaptation needs. While there is some skepticism about these figures (e.g., Agrawala and Fankhauser, 2008; Ranger et al., 2011), these figures are useful for advocacy purposes and estimating global investment requirements and related impacts. Thus far, estimates suggest high values of investment, associated with similarly high rates of returns and benefit-cost ratio. These figures generally include both evaluations of cost of damage and losses without investment and investment costs themselves. In the following paragraphs, we review these estimates.

Considering first the cost of damage in the absence of investment, the Stern Review (2006) estimated the value of failing to act would be equivalent to five percent of gross domestic product (GDP) annually, but this figure could rise to 20 percent, if earlier mitigation or adaptation actions are not taken. Stern adds that this effective five percent reduction in consumption globally should be viewed as "at a minimum, now and forever." This figure is supported by De Bandt et al. (2021) who find that an increase of a single degree (Celsius) in temperature results in decreased real GDP per capita annual growth rate of 1.13 percentage points. And calculations from Mogelgaard et al. (2018) suggest that if current trends continue to 2030, communities could lose 1 - 12 percent of their GDP because of climate change. With greater and/or unmitigated climate change, these losses could increase up to 200 percent as early as 2030. Bresch (2016) estimates that the current annualized damage of 1 to 12 percent of GDP resulting from climate change risks will rise to 19 percent of GDP by 2030.

There is some evidence that these losses can be mitigated cost effectively. Considering the value of adaptation investment, the World Resources Institute (2019) calculates cost-benefit ratios from 2:1 to 10:1 and suggests that investing $1.8 trillion globally in just five domains (strengthening early warning systems, making new infrastructure resilient, improving dryland agriculture and crop production, protecting mangroves, and making water resources management more resilient) could generate more than $7 trillion in net benefits. Similarly, looking only towards investments in low-income and developing countries, Hallegatte (2012) estimated that $1 billion of investment would result in benefits between $4 and $36 billion while the World Bank (2010) estimates the cost between 2010 and 2050 of adapting to a 2 degree (Celsius) warmer world (by 2050) is in the range of $75 billion to $100 billion per year. Bresch (2016) estimates that between 40 and 65 percent of the projected increases can be averted cost-effectively. However, Chapagain et al. (2020) find that the climate finance pledges (around $100 billion by 2020 at the time of the writing) are far short of the estimated global adaptation costs.

A subset of the headline figures focus specifically on Africa, due to projections that the continent will be most impacted by climate change in the coming decades. Figures considering the cost to Africa alone find tremendous costs of climate change impacts, on the scale of $10 billion to $90 billion by 2020 (Dinar et al., 2012). If actions are not taken to mitigate and address climate change in the coming years, Dinar et al. (2012) suggest that the scale of costs and required investments may increase to $50 to $100 billion by 2100. Mo Choi et al. (2020) estimate adaptation costs at $30 to $50 billion per year (which is approximately 2 - 3 percent of GDP) each year over the next decade (from 2020). The authors further assert that the estimated cost of investing in adaptation is far less than the projected costs of disaster relief in response to growing humanitarian need if no action were to be taken.

With respect to climate adaptation across Africa, the Tunis Roundtable (2010) contains several specific estimates. They project that the adaptation financing needs, as of 2010, ranged from $5 to $30 billion per year for the continent. While a large range exists for adaptation costs, the costs of climate change impacts are somewhat narrower: the projections suggest that each African economy is likely facing losses of one to two percent of gross domestic product (GDP) annually, approximately $10 to $20 billion annually (Van Aalst, et al., 2007).

Taken together, these topline figures and estimates of costs and benefits affirm the economic case for investing in climate adaptation both in Africa and globally, albeit at an extraordinary upfront cost with benefits accrued over time, including in the form of averted humanitarian need. These figures, the authors acknowledge, can be difficult to comprehend, particularly when simply presented as a figure, without consideration of methods, context, data, ranges, assumptions, or other essential details. To this end, in the next subsection, we investigate and synthesize the dominant methods in the literature of evaluating and projecting these estimated economic costs of climate adaptation.

*Methods, Assumptions, and Practice in Estimation*

In the past two decades, there has been an evolution in climate adaptation assessment. Beginning from what is called impact assessment or an impact pathways framework, estimation methods have become more complex, with additional assumptions and data requirements (Fankhauser and Soare, 2013). All models caution limitations with their conclusions, almost regardless of method. Stern (2006) notes that, in his exercise to quantify the cost of business as usual, there is a wider range of possible impacts of climate change than are possible to adequately quantify and thus to incorporate into the model. He notes further that the estimates are predicted on severe impacts, but because of the limitations of models, with respect to assumptions and data availability, these are likely underestimates. This, however, is still the case in much of the literature today, despite increasing data availability (e.g., Tröltzsch et al., 2016; World Bank, 2022).

*Cost Benefit Analysis*

As methods and practices have evolved, various different methods have been used for assessing the economics of adaptation (UNFCCC, 2009). Among these, cost-benefit analysis is the most commonplace. The literature is filled with papers that discuss how to implement these cost-benefit methods and approaches within various contexts (World Bank, 2010; Hallegatte et al., 2011; USAID, 2013; ECONADAPT, 2015; Watkiss, 2015; Rouillard et al., 2016; Markanday et al., 2021; Scussolini et al., n.d.).

Cost-benefit analyses are attractive as they give a clear, concise method for estimating the net present value of an investment, process, or other adaptation practice. Further, though there are significant data requirements, cost-benefit analysis is practicable and doable in most contexts, particularly when compared with other methods. There are also constructed flexibilities which are useful in examining the economics of adaptation. Consider Wilby and Dessai (2010), who explore the economic value of the same adaptation decision in the water sector using a variety of different methods, including scenario-led (or top down), vulnerability-based (or bottom up), or a combination of both. Their use of cost benefit analysis allows them to assess climate risk and adaptation in a more dynamic and flexible framework. In another example, Khan et al. (2012) investigates several case studies of disaster reduction interventions, investigating the use of cost-benefit analysis to evaluate individual projects versus portfolios of projects or programs. Finally, USAID (2013) identifies both quantitative and qualitative case studies of cost-benefit analysis. The authors find that there are important flexibilities in cost-benefit analysis (e.g., the incorporation of qualitative data) that can overcome often cited shortcomings with the method.

These flexibilities are sometimes not appreciated in much of the literature that criticizes cost-benefit analysis. The most interesting example of these flexibilities uses qualitative data or other methods to ensure the incorporation of non-market costs and benefits. For example, Cartwright et al. (2012), create a "people-centered" metric in order to assess adaptation effectiveness. The authors determine a population impact equivalent, in which the number of people benefitting from a particular intervention are identified, as well as the extent of that particular benefit for that population.

However, practices that take broader benefits like this into account are still uncommon. As such, cost-benefit analysis is often, rightfully, observed to be insufficiently flexible and to thus fail to capture dynamics that are important to calculating the full benefits and costs of climate adaptation processes (e.g., Cartwright et al., 2012; USAID, 2013). Some of this is due to discount rates, which may misestimate the value of various actions over time. Most notably, Parry et al. (2007) observes that most assessments of climate adaptation simply pair risks with an inventory of possible adaptation actions. The authors seem to suggest that these methods are often called cost-benefit analyses but are not actually cost-benefit analyses as would be done in standard economic practice. There are implications to using these methods: they result in estimates actually only covering a fraction of the actual costs, for example 30 to 50 percent of total costs in the health sector (Parry et al., 2007).

Even with its limitations, some argue for more literature and case studies using cost-benefit analysis for adaptation: cost-benefit analysis is still perhaps the best and most accessible method for determining whether an adaptation strategy is cost-effective and appropriate to undertake (Hallegatte et al., 2011; Markandy et al., 2019). Its dominance in the literature appears to be a reflection of that, though its dominance may simply reflect its ease of use and implementation. The preponderance of cost benefit analysis, though, is difficult to dispute. Abidoye (2021) is the most recent review of the literature. Abidoye (2021) cites limited literature about the economics of adaptation, including a table citing three papers. It is worth noting that these rely on cost-benefit analysis.

*Beyond Cost Benefit Analysis*

There is a smaller set of publications that emphasizes and uses methods in addition or as an alternative to cost-benefit analysis. For example, Lempert and Schlesigner (2000) promote exploratory models based in narrative scenario planning and quantitative decision analysis. Similarly, Lonsdale et al. (2008) develop adaptation solutions through a "process of dialogue". Both Bonjean and Downing (2011) and Cockburn (2000) emphasize the importance of the movement from understanding and studying adaptation as something static in the biophysical domain, to a more dynamic understanding of the decision processes, integrating social decision making. Alternatively, Rivadeneria and Carton (2022) discuss the role of integrated assessment models, as an approach for envisioning different scenarios, though they specifically focus on mitigation, rather than adaptation. Opportunities for extending these models to adaptation, however, are evident. Working to address some of the specific issues with cost-benefit analysis, Tröltzsch et al. (n.d.) highlights the use of cost-effectiveness analysis and multi-criteria analysis, in particular for use of non-monetary metrics. Finally, Bresch (2016) presents a new method: the Economics of Climate Adaptation (ECA), developed by a working group of the same name. The method proposes the pathway for decision makers to assess potential damages due to climate changes and then to determine which measures will be most cost effective in their implementation. The ECA develops a tool which permits decision makers to assess questions in a systematic way and to integrate adaptation "with economic development and sustainable growth". ECA offers an opportunity for minimizing the decision-making gap, but faces similar challenges to other methods associated with the amount of data required, uncertainty in estimates, and complexities in use.

Another robust and alternative method increasingly present in the literature on the economics of adaptation is that of real options analysis. These interesting studies include Dobes (2010), Abadie et al., (2017b), Dawson et al., (2018), Ginbo et al. (2020), Wreford et al. (2020), and Sims et al. (2021). A real options method still focuses, like cost-benefit analysis, on finding a net present value for an investment, but allows for a flexible approach to climate adaptation decisions. The real options approach requires significantly more data than standard cost-benefit analysis, however, and thus its use is likely limited, as many studies that employ cost-benefit analysis are already hampered by data limitations. But real options methods offer opportunities to handle uncertainty (Dawson et al., 2018). However, Kwakkel (2020) argues that real options analysis is not applicable or suitable for decision-making in climate adaptation due to profound issues with the way in which the method incorporates uncertainty.

In summary, though efforts exist to move beyond cost-benefit analysis, there is little consensus around alternatives to these methods, and most work that explores alternative methods focus on case studies in a single location and context. Despite documented and noted shortcomings with cost-benefit analysis, alternative methods often fail to fully address these shortcomings. Downing (2012) observes that, to some extent, this is by default: some work on the economics of adaptation fails to embrace and grapple with the complexity of climate change.[4] Much of this work relies on working from reference points, comparative statics, and climate impact scenarios, which fails to address known shortcomings with estimating the value of adaptation.

***Shortcomings and Limitations of Evidence in the Literature***

Beyond methodological limitations and related concerns, there is conversation in the literature on the economics of climate adaptation about opportunities for improvement and best practices and assumptions for estimation within those methods. The literature generally concludes - and we do not disagree - that while figures from the current literature are effective for advocacy purposes, many current approaches fall short of what is needed to inform policy and investment decision-making.

---

[4] Specifics of what is not included is explored in a subsequent section. This includes uncertainty, missing markets, distributional inequalities, and feedback loops.

In looking to improve, the literature has coalesced around a set of problems with current estimation, in particular around the assumptions required to make estimates and otherwise assess the value of climate adaptation. In this subsection we discuss the challenges of uncertainty, missing markets, distributional inequalities, and feedback loops in accurately and precisely measuring the value of climate adaptation.

*Uncertainty*

A broad consensus surrounds the difficulty in dealing with uncertainty in climate change and the potential for a range of climate futures. This challenge was noted as early as the 1970s, including by William Nordhaus who then referred to some climate predictions as "measurement without data" (Keen, 2020). Although climate science has dramatically advanced since that time, it is still not possible to predict the likelihood or magnitude of specific kinds of climate events in particular locations within specified time periods. This makes it extremely difficult to calculate the benefits associated with investments in adaptation in a particular sector or place. This challenge has been identified by Reynolds et al. (2007), UNFCCC (2009), Wilby and Dessai (2010), World Bank (2009), Stage (2010), De Bruin and Ansink (2011), Hallegatte et al. (2011), Agrawala et al. (2011), Fankhauser and Soare (2013), Watkiss et al. (2015), Watkiss (2015), Ginbo et al. (2020), and many others.

The economic adaptation literature is hardly the first to identify the difficulties in incorporating uncertainty into models, nor the first to fail to act on these challenges. But within the literature of the economics of adaptation, there are additional unique challenges to incorporating uncertainty. As Hallegatte et al. (2012) observe, we cannot eliminate deep uncertainties in the short-term (and probably not even in the long-term) in the climate space. The authors suggest that because of this, we must pursue estimates of uncertainty, even if they are imperfect. Montier et al. (2022) note that there are trade-offs when including uncertainty in estimates, often between accuracy and timing. But they note that it is necessary to include a "good enough" estimate.

But the issue is not simply uncertainty itself. Hallegatte et al. (2011) further note that the issue with insufficient incorporation of uncertainty goes beyond uncertainty about the trajectories of climate change. As adaptation is not necessarily a specific single action aimed at a single outcome, it can be difficult to capture the spillover outcomes of adaptive actions as well as the compounding economic effects, hence generating more uncertainty.[5] Further, climate uncertainty is unique, as its very trajectory depends on how quickly and in what way we collectively take action on the greenhouse gas emissions side of the equation. And so, this uncertainty limits the methods used by most studies estimating the value of investment in adaptation.

---

[5] One method for doing this is adaptation pathways. Though we do not discuss this at length, adaptation pathways address short- and long-term adaptation priorities together, accounting for social and economic context for adaptation, which shifts naturally over time. These pathways offer the opportunity to revisit and adjust strategies for adaptation, based on spillovers and changing circumstances (Marks et al., 2021).

Despite the challenges, there is work that includes and incorporates uncertainty and provides a model to do so in other contexts (Hallegatte et al., 2012; Hallegatte, 2013; Watkiss and Cimato, 2016; Abadie et al., 2017a; Wilby et al., 2021). Discounting has been noted as a potential method for addressing uncertainty (Stern, 2006; Chambwere et al, 2014; Trück et al., 2020 Wise et al., 2022) though evidence is still emerging about its application in practice.

Jafino et al. (2021) go further and argue that an emphasis on the differences across climate scenarios may result in bad adaptation policy advice. By focusing on climate impacts and their minimization over the achievement outcomes such as reduced inequality, inferior outcomes may result. The authors instead propose that work should focus on how policies influence the absolute level of metrics of interest (such as inequality) in scenarios with climate change, rather than focusing on how adaptation policies may attenuate or affect incremental climate impacts. Van Ginkel et al. (2020) propose that analysis should be based around tipping points, focusing on those drivers which affect step changes in impacts.

However, the uncertainty in estimates of climate change, and the uncertainty around the cost of climate change, has been used by policymakers to justify inaction. Lonsdale et al. (2008) observes that what has occurred with respect to adaptation is effectively "policy paralysis in response to what is a highly uncertain phenomena…". But, instead, we should follow the note of the Tunis Roundtable (2010) which observes that "The current state of knowledge is not good enough to provide firm projections. It is inappropriate to design adaptation strategies against a single future projection of modeled climate.", suggesting flexibility in adaptation in response to uncertainties.

Regardless of this challenge, it is essential to use the tools developed to generate better estimates of the economic costs of climate adaptation. Dessai et al. (2009) state that "society can (and must indeed) make effective adaptation decisions in the absence of accurate and precise climate predictions." And Abidoye (2021) notes: "uncertainty does not imply ignorance". As he suggests, the difficulty of incorporating uncertainty into models is still frequently used as an excuse for inaction relative to improving methods and models - or worse, to the real world, beyond the literature, in protecting life, property and ecosystems.

*Missing Markets (Going Beyond 'Economic' Value)*

A second issue highlighted throughout the literature, including Stern (2006), UNFCCC (2009), World Bank (2009), Watkiss (2015), Asplund and Hjerpe (2020), Bharadwaj et al. (2023), and others is the difficulty in estimating economic values for things that may have no "natural" dollar (e.g., economic) value. Economists identify this as a problem of missing and/or misaligned markets. Entities traded in a market can be valued with a common currency, and tools readily exist for their valuation (Fowler and Dunn, 2014). However, in this case, it's "what you don't count that counts'' as Keen (2020) writes. Thus, non-market valuation is a constant and often contentious issue in the literature in the economic case for climate adaptation. There is an extensive literature about this, outside of the adaptation literature, discussing how, why, and when to place a dollar value on various non-market resources such as the existence value or the value of biodiversity. Tröltzsch et al. (n.d.) present a synthesis of the available literature on non-monetary metrics employed in adaptation assessments, including for use in cost-effectiveness analysis and multi-criteria analysis.

Within the economic adaptation literature, there is some discussion of the shortcomings of non-market valuation, particularly in cost-benefit analysis. As Downing (2012) notes, it is difficult to price things that are truly irreplaceable: "Where the adaptation options involve market exchanges - such as the cost of land - adaptation falls within conventional decision frameworks. However, where adaptation requires solving socially contingent values - such as the loss of a culture associated with a vulnerable island - economic valuation is not easily achieved."

Many of the studies in sub-Saharan Africa exemplify the inappropriate cost estimates of climate change by using the average income of people on the continent in order to ascertain the value of damage from climate change. This might be the "correct" way for this to be done. However, it may result in undervaluing the damage, due to the prevalence of poverty on the continent and the fact that adaptive actions may create pathways out of hunger and poverty that are unaccounted for. This may also be the case in Kull et al. (2008), who show that benefits do not outweigh the costs of adaptation, perhaps due to the low monetary value assigned to individuals and property who benefit from the adaptation.

As Downing (2012) observes, despite the ease of doing so, "…it is not desirable to reduce all of the issues surrounding adaptation strategies and actions to a single dimension of monetary valuation." But much of the literature has done exactly that in the absence of methods for valuing things such as culture and psychosocial well-being.

*Distributional Inequalities*

Tying into the complexities of non-market valuation are the issues of distributional inequalities present in the economics of climate adaptation (Stern, 2006; UNFCC, 2009; Asplund and Hjerpe, 2020). Distributional inequalities refer to the way in which benefits and costs are spread unequally across populations, often with the poorest or most vulnerable people bearing the greatest "cost" and receiving the least "benefit". The economic calculations on the cost to vulnerable populations are generally unreliable. This is perhaps because climate change disproportionately affects poor and vulnerable groups (Watkiss and Cimato, 2016), but also because fewer data are generally available for less communities in the Global South. Thus, a key concern is whether and how these differences are accounted for, when assessing climate risks and adaptation processes. As Markanday et al. (2021) find, economic outcomes may change when equity concerns are included in the estimation.

Sovacool et al. (2015) identify a typology of the political economy of climate change adaptation in practice, which includes economic, political, ecological, and social dimensions through which distributional inequalities may occur and existing inequities may be exacerbated. Okuda and Kawaski (2022) note that poverty itself is a barrier to adaptation and so Fankhauser and Soare (2013) highlight the need to determine who should adapt and the barriers to that adaptation, which may be related to financial constraints and existing inequalities and inequities within cultures. Watkiss (2015) would classify these failures as governance and behavioral failures.

Uncertainty and feedback loops (discussed in the next subsection) amplify distributional inequalities, particularly with the acknowledgement that countries become less vulnerable (by some measures) as their economies grow (World Bank, 2010) and so distributional inequalities may attenuate. But assessing distributional inequalities offers one of the greatest unmet challenges for advancement in studies on the economics of climate adaptation (Fowler and Dunn, 2014).

Further, evidence suggests that investments in adaptation are made unequally, exacerbating inequities within the adaptation process. Ford et al. (2011) in their review of nearly 2,000 studies, find that there is limited adaptation? action related to vulnerable groups, despite discussion in the literature about the importance of making such investments. In some cases where investments are made they are made in a way that is not necessarily aligned with greatest need. This evidence is supported by Touboul (2021) who finds that there is a mismatch between countries' adaptation needs and their access to technologies necessary for adaptation programs and practices.

As with the case of uncertainty, tools exist to address distributional concerns. Hallegatte et al. (2011) acknowledge that distributional concerns are present in almost all projects that require investment and so there are tools, some inside and some outside, the adaptation space, that are designed to address these concerns and challenges. Using a different method, Downing et al. (2006) use vulnerability as a lens for exploring adaptation solutions to better capture social and economic vulnerabilities of vulnerable groups. Addressing the challenges of estimating distributional outcomes must be addressed in order to effectively value climate change and climate adaptation in relation to the commitment of USAID (2022) and others to include equity and inclusion as top priorities in their climate strategies.

*Feedback Loops*

Related to all the issues above, the problem of feedback loops and amplifying feedback (alternatively called spillover effects) have been long identified as a challenge in the literature (Stern, 2006; USAID, 2013). Kull et al. (2008) observe that feedbacks of climate impacts? into other sectors and domains are frequently omitted from studies, for example: "but it is not exact since it does not take into account the ways in which one expenditure may contribute to reducing several types of climate change harm."

Although there is discussion of feedback - or, at the very least, the need to adjust and accommodate possible feedback loops, these feedbacks are often considered to be multiplicative and unpredictable. Adger et al. (2007) note that adaptation does not always translate as expected (e.g., into reduced vulnerability). This is rarely addressed in the literature. One exception to this is Emolieva et al. (2016) who compare a global recursive-dynamic, partial-equilibrium model to explore spillovers and feedback loops, with stochastic and deterministic models. This exercise, though, requires a significant amount of data and modeling skill, as well as knowledge of a particular model platform (GLOBIOM) and so its applicability to other contexts may be limited.

Feedback loops are difficult to consider in some adaptation solutions, including insurance and risk sharing and public-private partnerships, as they further exacerbate uncertainty about the valuation of resources in the future (OECD, 2008; Green Climate Fund, 2019). Further, there are cascading impacts of climate change, which can cause both negative and positive feedback loops. In this situation, a single adaptation could solve multiple problems, but also potentially generate new challenges. These feedbacks and amplifying feedbacks within social and physical systems are challenging to incorporate into estimates of the economics of adaptation, but their omission likely underestimates the impact of any single adaptation action.

## 4. Additional Considerations and Limitations

In this section, we provide a final synthesis of the literature; in combination with the challenges noted above, this section aims to identify opportunities for future analyses, as well as opportunities for future reviews, which might serve the objective of better understanding the values of investments and incorporation of adaptation practices into policies and programs.

Put simply, it is a challenge to value the return on climate adaptation investments when current, process-focused approaches to measuring climate adaptation effectiveness (focused primarily on climate action and/or climate finance) do not accurately capture the impact of these investments in terms of whether (or not) communities and countries are actually better adapted to a changing climate as a result. This constitutes a major constraint for the more rigorous interrogation of the economic case for climate adaptation. As Cisse (2023) notes, USAID's pioneering work on resilience measurement in relation to near-term shock events provides an apt starting point. However, it too will need to evolve to address the uncertainty and time horizons associated with climate change.

In addition to addressing this major constraint, we highlight three more opportunities for future work. The first is the opportunity for causal studies, with a constructed or actual counterfactual. The second is the possibility for a more clarified and united language in the climate adaptation space. The third is deliberate collection of data that allows for adaptive learning relative to the costs and benefits of investments in adaptation. We discuss these each in turn and how further work might elucidate possible insights from them.

At present, there are very few causally-analyzed, counterfactual-implemented studies, whether through randomization, natural experiments, quasi-experimental approaches, or other causal frameworks in the literature. While more complicated than current methods used to assess the value of adaptation, there are opportunities for well-structured, thoughtful studies to shed light on the impacts of adaptation being realized. This would require an analytic framework which differentiates those who adopt adaptation strategies or engage in adaptation processes, compared with those who do not, in terms of the impact on current and future well-being, even if the impact on future well-being has to be modeled against a range of possible climate futures. This presents an opportunity for future work, building on existing work focused on measuring resilience in relation to near-term shocks by USAID and others.

One major caveat necessary here is that the time frames associated with climate change impacts pose a significant challenge to identifying causal impacts. Trück et al. (2020) observe that differential time scales may disproportionately deflate the risk involved and/or artificially or incorrectly inflate costs. But, this suggests a need to focus on both near-term and longer-term impacts as suggested above, rather than to altogether abandon the prospect. There may also be opportunities to identify early signals of longer-term impacts. To date, there are few to no credible counterfactual studies for adaptation measures with respect to medium- and long-term climate change impacts.

Another opportunity that emerges from the assessment of the literature is the need for a clarification and (as possible) unification of language in this space. It is a limitation of this work that we may have potentially "missed" from studies with adaptation and climate resilience as the key search terms. There are myriad terms in the literature that might investigate the same concept, but use different language to do so, due to the dispersed nature and varied backgrounds of communities working on this issue. But the consequence is clear: activities that are effectively adaptation but are not identified as such fail to be captured.[6] Consider Abidoye (2021), which highlights several studies that discuss adaptation practices, such as the use of biochar (Dickinson et al., 2014). Biochar (used as a soil amendment) is not exclusively a practice undertaken for the purpose of mitigation and adaptation (as Adger et al. (2007) note, very few actions are). The use of biochar is often used as a mitigation strategy, while organic soil amendments support retained soil moisture and thus can also be an adaptive strategy. While Dickinson et al. (2014) use an adaptation framework, there are numerous other studies that do not. To illustrate this, consider another study (by one of the authors of this review), Josephson & Ricker-Gilbert (2020). In this, the authors consider adoption and dis-adoption of sorghum, a drought tolerant crop, in Zimbabwe. The framing of the paper is one of subsistence farming in missing markets, but it could also be one of adaptation, in which farmers are adapting to a changing climate, in the context highlighted in the paper. A unified language in the climate adaptation space and/or an expanded analysis of what qualifies as a climate adaptation may yield further insight about the status of adaptation and the economics of its undertaking.

There are likely many more studies, in particular in the economics literature, which cover these sorts of adaptive actions and adaptation processes, but do not use the language of adaptation or climate resilience.[7] In fact, these studies might begin to address some of the limitations highlighted here, including the lack of causal research and work with counterfactuals. While a more expansive look beyond these key terms is outside the scope of this review, it would be a fruitful exercise for better understanding alternative ways for estimation of benefits and costs, as well as averted losses

The final gap and opportunity demonstrated by the existing literature is a lack of data at multiple scales and over multiple time frames. To address this, data, simulations, and/or estimates are often developed for unique scenarios and in different studies, which leads to disparities in conclusions and a general lack of cohesion in projections. Long-term efforts to collect appropriate data and metrics across geographies and time frames that go beyond typical project lifecycles would allow for adaptive learning relative to the costs and benefits of investments in adaptation. This could help to resolve these challenges and to create new opportunities for systematic, comparable analysis of the economics of climate adaptation.

.

---

[6] Note that the inverse is also true in that attempts to recast or greenwash many activities as climate-smart may cloud estimates of the economic value of adaptation actions.

[7] This may be emblematic of a further issue in which disciplines fail to communicate with one another, even when ostensibly doing interdisciplinary work (e.g., as noted by Rivadeneria and Carton (2022) and Bharadjwaj et al., (2023)).

## 5. Conclusion

In this paper, we have worked to synthesize and understand the current literature on the economic case for climate adaptation, with the objective of assessing the value (economic and otherwise) of climate change adaptation, as well as the strength of the methods and evidence behind it. The literature on the economics of climate adaptation, while extensive, is limited due to a variety of issues. But, even with these limitations, there is abundant evidence from case studies in various applications of adaptation and related estimates of costs and benefits. There is a useful and practical literature about implementation and practice of various methods and tools, designed specifically and created for practitioners. This includes World Bank (2010), Hallegatte et al. (2011), USAID (2013), ECONADAPT (2015), Watkiss (2015), Rouillard et al. (2016), Bouley et al. (2018), Mogelgaard et al. (2018), and Scussolini et al. (n.d.). There remain opportunities - even an imperative - to learn from and to extend this literature, better address the limitations highlighted within it and improve methodologies for future estimation, including by advancing and improving the measurement of climate adaptation itself. Doing so would better enable those on the front lines of the climate crisis to make evidence-based investment decisions against the backdrop of constrained resources for doing so.

**References**

Abadie, L.M., I. Galarraga, E. Sainz de Muierta. (2017a). Understanding risks in the light of uncertainty: low-probability, high-impact coastal events in cities. *Environmental Research Letters* 12: 014017.

Abadie, L.M., E. Sainz de Muierta, I. Galarraga. (2017b). Investing in adaptation: Flood risk and real option application to Bilbao. *Environmental Modelling and Software* 95: 76 - 89.

Abadie, L.M. (2018). Sea level damage risk with probabilistic weighting of IPCC scenarios: An application to major coastal cities. *Journal of Cleaner Production* 175 (20): 582 - 598.

Abidoye, B.O. (2021). Economics of climate change adaptation. Oxford Research Encyclopedias. Environmental Science.

AdaptNow. (2019). Adapt Now: A global call for leadership on climate resilience. Global Commission on Adaptation.

Asplund, T., M. Hjerpe. (2020). Project coordinators' views on climate adaptation costs and benefits - justice implications. *Local Environment: The International Journal of Justice and Sustainability* 25 (2): 114 - 129.

Adger, W.N., S. Agrawala, M.N. Q. Mirza. (2007). Assessment of adaptation practices, options, constraints, and capacity. Climate Change 2007: Impacts, Adaptation, and Vulnerability. Contribution of Working Group II to the Fourth Assessment Report of the Intergovernmental Panel on Climate Change, M.L. Parry, O.F. Canziani, J.P. Palutikof, P.J. van der Liden, C.E. Hanson, eds., Cambridge University Press, Cambridge, UK. 717 - 743.

Agrawala, S., S. Fankhauser. (2008). Economic aspects of adaptation to climate change. Costs, benefits and policy instruments. OECD, Paris.

Agrawala, S., F. Bosello, C. Carrao, E. de Cian, E. Lanzi. (2011). Adapting to climate change: Costs, benefits, and modeling approaches. *International Review of Environmental and Resource Economics* 5: 245 - 284.

Bachner, G., B. Bednar-Friedl, N. Knittel. (2019). How does climate change adaptation affect public budgets? Development of an assessment framework and a demonstration for Austria. *Migration and Adaptation Strategies for Global Change* 24: 1325 - 1341.

Banhalmi,-Zakar, Z., R. Hales. (2016). Guidance on how to build a business case for climate change adaptation: Lessons from coastal Australia. CoastAdapt, National Climate Change Adaptation Research Facility: Gold Coast.

Bharadwaj, R., Mitchell, T., Huq, S. (2023). Non-economic loss and damage: closing the knowledge gap. IIED, London https://www.iied.org/21311iied

Blyth, N. (n.d.). Climate change adaptation building the business case: Guidance for environment and sustainability practitioners. Institute of Environmental Management and Assessment (IEMA). Dept. for Environment Food and Rural Affairs.

Box, G.E.P. (1976). Science and statistics. *Journal of the American Statistical Association* 71 (356): 791 - 799.

Bonjean, M., T.E. Downing. (2011). Adaptation space: Use case builder. Oxford: Global Climate Change Adaptation Partnership.

Bouley, T., K.L. Ebi, A. Midgley, J. Shumake-Guillemot, C.D. Golden. (2018). Methodological guidance: climate change and health diagnostic: A country-based approach for assessing risks and investing in climate-smart health systems. World Bank Group, World Health Organization: Investing in Climate Change and Health Series.

Bresch, D.N. (2016). Shaping Climate Resilient Development: Economics of Climate Adaptation. *Climate Change Adaptation Strategies–An Upstream-downstream Perspective*: 241-254.

Bueb, B., J. Tröltzsch, D. Reichwein, C. Oldenburg, F. Favero. (2021). Case Studies of Sustainable Adaptation Pathways: Appendix to the Paper 'Towards Sustainable Adaptation Pathways: A concept for integrative actions to achieve the 2030 Agenda, Paris Agreement and Sendai Framework'. On behalf of the German Environmental Agency: Ressortforschungsplan of the Federal Ministry for the Environment, Nature Conservation and Nuclear Safety. Project No. FKZ 3719181050.

Carter, T., M. Parry, H. Harasawa, S. Nishioka. (1994). IPCC technical guidelines for assessing change impacts and adaptation. Dept. of Geography, University College London, London.

Carter, T. R.N. Jones, X. Lu, et al. (2007). New assessment methods and the characterisation of future conditions. In: Parry M. et al. (eds). Climate Change 2007: Impacts, adaptation, and vulnerability. Contribution of Working Group II to the Fourth Assessment Report of the Intergovernmental Panel on Climate Change.

Cartwright, A., J. Blignaut, M. De Wit, K. Goldberg, M. Mander, S. O'Donoghue, D. Roberts. (2012). Economics of climate change adaptation at the local scale under conditions of uncertainty and resource constraints: The case of Durban, South Africa. Environment and Urbanization 25 (1): 139 - 156.

Chapagain, D., F. Baarsch, M. Schaeffer, S. D'haen. (2020). Climate change adaptation costs in developing countries: insights from existing estimates. *Climate and Development* 12 (10): 934 - 942.

Chu, E., A. Brown, K. Michael, J. Du, S. Lwasa, A. Mahendra. (2019). Unlocking the potential for transformative climate adaptation in cities. Washington D.C. Global Commission on Adaptation and World Resources Institute. Rosenzweig, C., W. Solecki, P. Romero-Lankao, S. Mehrotra, S. Dhakal, S. Ali Ibrahim, eds. 2018. *Climate Change and Cities: Second Assessment Report of the Urban Climate Change Research Network (ARC3 2)*. Cambridge: Cambridge University Press.

Cissé, J.D. (2023). Opinion: Defining climate adaptation success is possible - and urgent. DevEx. https://www.devex.com/news/opinion-defining-climate-adaptation-success-is-possible-and-urgent-106412

Chambwera, M., G. Heal, C. Dubeau, S. Hallegate, L. Leclerc, A. Markandya, B.A. McCarl, R. Mechler, J.E. Neumann. (2014) Ch. 17: Economics of Adaptation. *In: Climate Change 2014: Impacts, Adaptation, and Vulnerability. Part A: Global and Sectoral Aspects.* Contribution of Working Group II to the Fifth Assessment Report of the Intergovernmental Panel on Climate Change.

Cockburn, A. (2000). Writing effective use cases, 270. Boston, MA: Addison-Wesley.

Daidone, S., B. Davis, S. Handa, P. Winters. (2019). The household and individual-level productive impacts of cash transfer programs in Sub-Saharan Africa. *American Journal of Agricultural Economics* 101 (5): 1401 - 1431.

Dawson, D.A., A. Hunt, J. Shaw, W.R. Gehrels. (2018). The economic value of climate information in adaptation decisions: Learning in sea-level rise and coastal infrastructure context. *Ecological Economics* 150: 1 - 10.

De Bandt, O., L. Jacolin, L. Thibault. (2021). Climate change in developing countries: Global warming effects, transmission channels and adaptation policies. Banque de France, Eurosystème. Working Paper 822.

De Bruin, K., E. Ansink. (2011). Investment in flood protection measures under climate change uncertainty. *Climate Change Economics* 2(4): 321 - 339.

Dessai, S., M. Hulme, R. Lempert, R. Piekle. (2009). Do we need better predictions to adapt to a changing climate? Eos Trans Am Geophys Union. 90: 111.

Dickinson, D., L. Balduccio, J. Buysse, F. Ronsse, G. van Huylenbroeck, W. Prins. (2014). Cost-benefit analysis of using biochar to improve cereals agriculture. *GCB Bioenergy* 7 (4): 1 - 15.

Dinar, A., R. Hassan, R. Mendelsohn, J. Benhin (2012). Climate change and agriculture in Africa: Impact assessment and adaptation strategies. EarthScan.

Dobes, L. (2010). Notes on applying 'real options' to climate change adaptation measures, with examples from Vietnam. CCEP Working Paper 7.10. Centre for Climate Economics and Policy, Crawford School of Economics and Government. Australian National University: Canberra.

Downing, T.E., J. Aerts, J. Soussan, O. Barthelemy, S. Bharwani, C. Ionescu, J. Hinkel, R.J.T. Klein, L. Mata, N. Martin, et al. (2006). Integrating social vulnerability into water management. SEI Working Paper and Newater Working Paper No. 4. Oxford: Stockholm Environment Institute.

Downing, T.E. (2012). Views of the frontiers in climate change adaptation economics. *WIREs Clim Change* 3: 161 - 170.

Eckersley, P., K. England, L. Ferry. (2018). Sustainable development in cities: collaborating to improve urban climate resilience and develop the business case for adaptation. *Public Money & Management* 38.5: 335-344.

ECONADAPT. (n.d.). Costs and Benefits of Adaptation. European Union Seventh Framework Programme: 2 - 7.

Emolieva, T., A. Biewald, E. Boere, P. Havlik. (2016). Overview report on major uncertainties related to climate impacts and socioeconomic costs, and policy recommendations related to the effectiveness of adaptation options. ECONADAPT. Deliverable 7.3.

Espinet, X., A. Schweikert, P. Chinowsky. (2015). Robust prioritization framework for transport infrastructure adaptation investments under uncertainty of climate change. *ASCE-ASME Journal of Risk and Uncertainty in Engineering Systems, Part A: Civil Engineering 3* (10): 101.161.

Fankhauser, S., R. Soare. (2013). An economic approach to adaptation: illustrations from Europe. *Climatic Change* 118: 367 - 379.

Filho, W.L., A-L. Balogun, O.E. Olayide, U.M. Azeiteiro, D.Y. Ayal, P.D. Chavez Muñoz, G.J. Nagy, P. Bynoe, O. Gouge, N.Y. Toamukum, M. Saroar, C. Li. (2019). Assessing the impacts of climate change in cities and their adaptive capacity: Towards transformative approaches to climate change adaptation and poverty reduction in urban areas in a set of developing countries. *Science of the Total Environment* 692: 1175 - 1190.

Ford, J.D., L. Berrang-Ford, J. Paterson. (2011). A systematic review of observed climate change adaptation in developing countries. Climatic Change 106: 327 - 336.

Fowler, Ben, and Elizabeth Dunn. (2014). Evaluating Systems and Systemic Change for Inclusive Market Development: Literature Review and Synthesis. Leveraging Economic Opportunities: Report 3. United States Agency for International Development (USAID).

Gao, S., W. Zhai. (2023). Assessing adaptation planning strategies of interconnected infrastructure under sea-level rise by economic analysis. *Frontiers of Architectural Research*: In press.

García Sánchez, F., W.D. Solecki, C.B. Batalla. (2018). Climate change adaptation in Europe and the United States: A comparative approach to urban green spaces in Bilbao and New York City. *Land Use Policy* 79: 164 - 173.

Ginbo, T., L. Di Corato, R. Hoffmann. (2020). Investing in climate change adaptation and mitigation: A methodological review of real-options studies. *Ambio* 50: 229-241.

Green Climate Fund. (2019). Adaptation: accelerating action towards a climate resilient future. Green Climate Fund, Working Paper No. 1.

Hale, R., Z. Banhalmi-Zakar, T. Sarker, A. Lo, A. Chai, E. Whittlesea, C. Fleming, K. Kelly, M. Bun. (2016). Guidance on how to build a business case for climate change adaptation. National Climate Change Adaptation Research Facility, Gold Coast.

Hallegatte, S. (2009). Strategies to adapt to an uncertain climate change. Global Environmental Change 19 (2): 240 - 247.

Hallegatte, S., F. Henriet, J. Coffee-Morlot. (2011b). The economics of climate change impacts and policy benefits at city scale. Climate Change 104 (1): 51 - 87.

Hallegatte, S., A. Shah, R. Lempert, C. Brown, S. Gill. (2012). Investment decision making under deep uncertainty: Application to climate change. Policy Research Working Paper 6193. World Bank, D.C.

Hallegatte, S. (2013). Challenges ahead: Risk management and cost-benefit analysis in a climate change context. In: D. Guha-Sapir, I. Santos (eds.). The Economic Impact of Natural Disaster. London: EarthScan. 107 - 127.

Hallegatte, S., C. Green, R.J. Nicholls, J .Corfee-Morlot. (2013). Future flood losses in major coastal cities. Nature Climate Change 3: 802 – 806

Hallegatte, S., F. Lecocq, C. de Perthuis. (2011a). Designing climate change adaptation policies: An economic framework. World Bank Policy Research Working Paper 5568.

Hallegatte, S. (2012). A cost effective solution to reduce disaster losses in developing countries: Hydro-meteorological services, early warning, and evaluation. World Bank Policy Research Working Paper 6058.

He, L., G. Li, K. Li, Y. Zhang, T. Guo. (2020). Damage of extreme water levels and the adaptation cost of dikes in the Pearl River Delta. *Journal of Water and Climate Change* 11 (3): 829 - 838.

Hughes, G., P. Chinowsky, K. Strzepek (2010). The costs of adapting to climate change for infrastructure. Development and Climate Change Discussion Paper. No. 2. World Bank: Washington, D.C.

Hurley, T.M., P.G. Pardey, X. Rao, R. Andrade. (2016). Returns to food and agriculture R&D investments worldwide, 1958-2015. InSTePP Brief. St. Paul, MN: International Science and Technology Practice & Policy Center.

Jafino, B.A., S. Hallegatte, J. Rozenberg. (2021) Focusing on differences across scenarios could lead to bad adaptation policy advice. *Nature Climate Change* 11.5: 394-396.

Jenal, M. (2019). Measuring Systemic Change in Market Systems Development: A Stock Taking. USAID. Monitoring and Evaluation Support for Collaborative Learning and Adapting (MESCLA) Activity.

Josephson, A., J. Ricker-Gilbert. (2020). Preferences and crop choice during Zimbabwe's macroeconomic crisis. *African Journal of Agricultural and Resource Economics* 15 (3): 260 - 287.

Keen, S. (2022). The appallingly bad neoclassical economics of climate change. *Economics and Climate Emergency*. *Globalizations*: 79-107.

Khan, F., M. Moench, S. Orleans Reed, A. Dixit, S. Shrestha, K. Dixit. (2012). Understanding the costs and benefits of disaster risk reduction under changing climate conditions: Case study results and underlying principles. March. ISET-International: Bangkok.

Kull, D., P. Singh, S. Copd, S. Wajih, and the Risk to Resilience Study Team. (2008). Evaluating costs and benefits of flood reduction under changing climatic conditions: Case of the Rohini River Basin, India. From Risk to Resilience Working Paper No. 4. M. Moench, E. Caspari, A. Pokhrel (eds.). ISET-Nepal and ProVention: Kathmandu, Nepal.

Kuruppu, N.D., S. Bee, C. Schaer. (2018). Developing the business case for adaptation in agriculture: case studies from the Adaptation Mitigation Readiness Project. Private-sector action in adaptation: *Perspectives on the role of micro, small and medium size enterprises*. UNEP DTU Partnership:. 51-62.

Kwakkel, J.H. (2020). Is real options analysis fit for purpose in supporting climate adaptation planning and decision-making?. *WIREs: Climate Change* 11.3: e638.

Lempert, R.J., M.E. Schlesigner. (2000). Robust strategies for abating climate change. Climatic Change 45 (3-4): 387 - 401.

Lonsdale, K.G., T.E. Downing, R.J. Nicholls, D. Parker, A.T. Vafeidis, R. Dawson, J. Hall. (2008). Plausible responses to the threat of rapid sea-level rise in the Thames estuary. Climate Change 91: 145 - 169.

Markanday, A., I. Galarraga, A. Markandya. (2019). A critical review of cost-benefit analysis for climate change adaptation in cities. *Climate Change Economics* 10.04: 1950014.

Markanday, A., A. Markandya, E. Sainz de Murieta, I. Galarraga. (2021). Accounting for the effects of employment, equity, and risk aversion in cost-benefit analysis: An application to an adaptation project. *Journal of Cost Benefit Analysis* 12 (2): 313 - 334.

Marks, M., J. Liu, P. Krans. (2021). Ramping-up climate adaptation through a "pathways" approach. ICF: Insights and Environment. Accessed 30 January 2024. https://www.icf.com/insights/environment/climate-adaptation-pathways-approach.

Mendelsohn, R. (2012). The economics of adaptation to climate change in developing countries. *Climate Change Economics* 3 (2): 1250006.

Mo Choi, S., M. Coelho, E.P. Endengle, W. Guo, K. Kalonji, A. Lagerborg, J. Li, G. Melina, E. Mensah, A. Thomas, M. Wang, J. Yago, G. Zinabou. (2020). Adapting to climate change in sub-Saharan Africa. Regional Economic Outlook: Sub-Saharan Africa. International Monetary Fund.

Mogelgaard, K., A. Dinshaw, N. Ginoya, M. Gutiérrez, P. Preethan, J. Waslander. (2018). From planning to action: Mainstreaming climate change adaptation into development. World Resources Institute. Working Paper. http://www.wri.org/publication/climate-planning-to-action.

Montier, E., L. Wrongärtner, S. Klassen. (2022). The potential for anticipatory action and disaster risk finance: Guiding the setting of humanitarian targets. The Potential for Anticipatory Action and Disaster Risk Finance. DI Start Network.

Ng'ang'a, S.K., V. Miller, G. Essegbey, N. Karbo, V. Ansah, D. Nautsukpo, S. Kingsley, E. Girvetz. (2017). Cost and benefit analysis for climate-smart agriculture (CSA) practices in the coastal savannah agro-ecological zone (AEZ) of Ghana. International Center for Tropical Agriculture (CIAT) Working Paper. CIAT Working Paper 132.

Nurmi, V., K. Pili-Sihvola, H. Gregow, A. Perrels. (2019). Overadaptation to climate change? The case of the 2013 Finnish electricity market act. *Economics of Disasters and Climate Change* 3: 161 - 190.

OECD (2008). Economic aspects of adaptation of climate change: Costs, benefits, and policy instruments, OECD Publishing Paris.

Okuda, K., A. Kawasaki. (2022). Effects of disaster risk reduction on socio-economic development and poverty reduction. *International Journal of Disaster Risk Reduction* 80: 103241.

Ozment, S., R. Feltran-Barbieri, P. Hamel, E. Gray, J.B. Ribeiro, S.R. Barreto., A. Padovezi, T.P. Valente. (2018). *Natural infrastructure in Sao Paulo's water system.* Washington D.C. World Resources Institute.

Parry, M., T. Carter. (1998). Climate impact and adaptation assessment: A guide to the IPCC approach. Earthscan, London.

Parry, M, et al. (2007). Climate change 2007: Impacts, adaptation, and vulnerability. Contribution of working Group II to the Fourth Assessment Report of the Intergovernmental Panel on Climate Change.

Pigato, M. (2019) Fiscal policies for development and climate action. Ed. M. Pigato. Washington, DC: World Bank Group.

Ranger, N., S. Hallegatte, S. Bhattacharya, M. Bachu, S. Priya, K. Dhore, F. Rafique, P. Mathur, N. Naville, F. Henriet, C. Herweijer, S. Phoit, J. Coffee-Morlot. (2011). A preliminary assessment of the potential impact of climate change on flood risk in Mumbai. Climatic Change. 104 (1): 139 - 167.

Reynolds, J.F., D.M Stafford Smith, E.F. Lambin, B.L. Turner II, M. Morimore, S.P.J. Batterbury, T.E. Downing, H. Dowlatabadi, R.J. Fernadez, J.E. Herrick, et al. (2007). Global desertification: building a science for dryland development. Science 316: 847 - 851.

Rivadeneira, Natalia Rubiano, and Wim Carton. (2022). (In) justice in modelled climate futures: A review of integrated assessment modelling critiques through a justice lens. *Energy Research & Social Science* 92: 102781.

Rouillard, J., J. Troltzsch, J. Tarpey, M. Lago, P. Watkiss, A. Hunt, F. Bosello, T. Ermolieva, C. Goddess, R. Mechler, R. Parrado, E. Sainz de Murieta, P. Scussolin. (2016). The economic analysis of climate adaptation: Insights for policy-makers. ECONADAPT.

Ryan, P.C., M.G. Stewart. (2017). Cost-benefit analysis of climate change adaptation for power pole networks. *Climatic Change* 143: 519-533.

Sahay, S. (2019). Adaptation to health outcomes of climate change and variability at the city level: An empirical decision support tool. *Sustainable Cities and Society*: 47: 101512.

Scussolini, P., O Kuik, A. Aerts, T. Veldkamp, P. Hudson, E Sainz de Murieta, I. Galarraga, K. Kaprova, J. Melichar, M. Lago, J. Rouillard, J. Troeltzsch, A. Hunt, M. Skourtos, C. Goddess, O.B. Christensen. (n.d.). The economic appraisal of adaptation investments under uncertainties: Policy recommendations, lessons learnt, and guidance. ECONADAPT.

Simpkins, G. (2021). The costs of infrastructure adaptation. Nature Reviews Earth and Environment 2 (661).

Sims, C., S.E. Null, J. Medellin-Azuara, A. Odame. (2021). Hurry up or wait: Are private investments in climate change adaptation delayed? *Climate Change Economics* 12 (4): 2150012.

Smit, B., O. Pilifosova (2001). Adaptation to climate change in the context of sustainable development and equity. IPCC, Ch. 18. Working Group II: Impacts, Adaptation and Vulnerability.

Smith, D.M., J.H. Matthews, L. Bharati, E. Borgomeo, M. McCartheny, A. Mauroner, A. Nicol, D. Rodriguez, C. Sadoff, D. Suhardiman, I. Timboe, G. Amarnath, N. Anisha. (2019). Adaptation's thirst: Accelerating the convergence of water and climate action. Background Paper prepared for the 2019 report of the Global Commission on Adaptation, Rotterdam and Washington DC. https://cgspace.cgiar.org/bitstream/handle/10568/106030/adaptations-thirst-gca-background-paper.pdf

Sovacool, B.K., B. Linner, M.E. Goodsite. (2015). The political economy of climate adaptation. *Nature Climate Change* 5: 616 - 618.

Stage, J. (2010). Economic valuation of climate change adaptation in developing countries. *Annals of the New York Academy of Sciences* 1185.1: 150-163.

Stern, N.H. (2006). *The economics of climate change: the Stern report.* Cambridge University Press, Cambridge, U.K. [Chapters 2, 6, 18]

The Adaptation Committee. (n.d.). The business case for adaptation.

The Nature Conservancy and C40 Cities. (2016). *A global analysis of the role of urban trees in addressing particulate matter pollution and extreme heat.*

Touboul, S. (2021). Technological innovation and adaptation to climate change. Thèse de doctorat en Economie. Soutenue le 03-12-2021 à l'Université Paris science et lettres dans le cadre de Ecole doctorale SDOSE (Paris), en partenariat avec Centre d'économie industrielle (Paris) et de Ecole nationale supérieure des mines (Paris). https://www.theses.fr/2021UPSLM054.

Tröltzsch, J., J. Rouillard, M. Lago, M. Hasenheit. (2016). ECONADAPT Toolbox: Data repository. ECONADAPT deliverable 10.5.

Tröltzsch, J., M. Lago, J. Rouillard, E. Lukat, K. Abhold, S. Beyer, M. Hinzmann, P. Watkiss, A. Hunt. (n.d.) Reviewing the use of non-monetary metrics/weights for use in decision-making of adaptation to climate change projects. Presentation: ECONADAPT.

Trück, S., C. Truong, T. Keighley, F. Liu, S. Mathew. (2020). Climate change and extreme events: Risk assessment of adaptation in Sydney. *Industry and Higher Education* 586: 101 - 130.

Tunis Roundtable. (2010) Economics of Adaptation. AdaptCost project: United Nations Environment Programme.

UNFCCC (2009). Potential costs and benefits of adaptation options: A review of existing literature. Technical Paper. Markandya, A. and Watkiss, P.

USAID. (2013). Methods for economic analysis of climate change adaptation interventions: African and Latin American Resilience to Climate Change.

Van Aalst, M., M. Hellmuth, D. Ponzi. (2007). Come rain or shine: Integrating climate risk management into African Development Bank Operations. Working Paper No. 89. African Development Bank, Tunis.

Van Ginkel, K.C.H., W.J. Wouter Botzen, M. Haasnoot, G. Bachner, K.W. Steininger, J. Hinkel, P. Watkiss, E. Boere, A. Jeuken, E. Sainz de Murieta, F. Bosello. (2020). Climate change induced socio-economic tipping points: review and stakeholder consultation for policy relevant research. *Environmental Research Letters* 15.2: 023001.

Watkiss, P. (2015). A review of the economics of adaptation and climate-resilient development. Centre for Climate Change Economics and Policy Working Paper 231, Grantham Research Institute on Climate Change and the Environment Working Paper 205.

Watkiss, P., A. Hunt, W. Blyth, J. Dyszynski. (2015). The use of new economic decision support tools for adaptation assessment: A review of the methods and applications, towards guidance on applicability. *Climatic Change* 132: 401 - 316.

Watkiss, P., F. Cimato. (2016). The economics of adaptation and climate-resilient development: Lessons from projects for key adaptation challenges. Centre for Climate Change Economics and Policy, Working Paper 265. Grantham Research Institute on Climate Change and the Environment, Working Paper 235.

Wilby, R.L., S. Dessai (2010). Robust adaptation to climate change. Weather 65 (7): DOI: 10.1002/wea.543.

Wilby, R.L., X. Lu, P. Watkiss, C.A. Rodgers. (2021). Towards pragmatism in climate risk analysis and adaptation. *Water Policy* 23 (S1): 10 - 30.

Wise, R.M., T. Capon, B.B. Lin, M. Stafford-Smith. (2022). Pragmatic cost–benefit analysis for infrastructure resilience. *Nature Climate Change* 12.10: 881-883.

Wimberly, M. C., and D. M. Nekorchuk. (2021). Malaria early warning in Ethiopia: a roadmap for scaling to the national level. *Washington, DC: United States Agency for International Development*.

World Bank. (2009). Evaluating adaptation via Economic analysis. Guidance Note 7, Appendix 12.

World Bank (2010). The costs to developing countries of adapting to climate change: New methods and estimates. The Global Report of the Economics of Adaptation to Climate Change Study. Synthesis Report. World Bank, Washington D.C.

World Bank (2022). Climate and development: An agenda for action. Emerging insights from World Bank Group 2021-22 Country Climate and Development Reports.

World Resources Institute. 2019. Estimating the economic benefits of climate adaptation investments. Technical Paper.

Wreford, A., R. Dittrich, T.D. van der Pol. The added value of real options analysis for climate change adaptation. *WIREs: Climate Change* 11.3 (2020): e642.

Yaron, G., D. Wilson. (2020). Estimating the economic returns to community-level interventions that build resilience to flooding. *Journal of Flood Risk Management* 13.4: e12662.